\documentclass[twocolumn,epsf,aps,amsmath,prl]{revtex4}
\usepackage{epsfig,psfrag}
\begin{document}
%\draft

\title{Revisiting the Hanbury Brown-Twiss set-up for fractional statistics}

\author{Smitha Vishveshwara}
\address{Department of Physics, University of Illinois
at Urbana-Champaign, Urbana, Illinois 61801-3080}

\date{\today}
%\maketitle

 \begin{abstract}
 The Hanbury Brown-Twiss experiment has proved to be an effective means
 of probing statistics of particles. Here, in a set-up involving  
edge-state quasiparticles in a fractional quantum Hall system, we show
 that a variant of the experiment composed of two sources and two sinks
can be used to unearth fractional statistics. We find a clear cut
signature of the statistics in the equal-time current-current correlation
function for quasiparticle currents emerging from the two sources and
collected at the  sinks.

\end{abstract}


\maketitle

%\begin{multicols}{2}
\narrowtext
The statistics of indistinguishable particles is manifested in
 the fate of the common wavefunction for two particles 
located at positions $r_1$ and $r_2$ under exchange:
\begin{equation}
\Psi(r_1, r_2, r_3,\ldots, r_n)
 = e^{\pm i \pi \nu} \Psi(r_2, r_1, r_3,\ldots, r_n),
\label{wvfn}
\end{equation}
where, in general, other particles may be present at positions 
`$r_i$', $i\neq 1,2$. 
The values $\nu=2n$ and $\nu=2n+1$ for integers `$n$'  correspond 
to the familiar instances of bosonic and fermionic statistics, 
respectively. In two dimensions, where the concept of exchange 
can be unambiguously defined, 
$\nu$ can assume fractional values corresponding to anyonic statistics. 
A landmark example of this phenomenon occurs for Laughlin 
states\cite{Mr.Bob} in the fractional quantum Hall (FQH) set-up. In this 
system, the anyonic nature of quasiparticle/quasihole excitations has been 
demonstrated\cite{Halperin} and, in particular, the gaining of the 
phase factor $e^{\pm i \pi \nu}$ by quasiparticles\cite{Jain1} 
and quasiholes\cite{Arovas,Wenrev,Eduardo} under exchange, 
where $\nu$ is the filling fraction.
Of late, a variety of novel proposals
for testing the statistics of edge-state quasiparticles in Laughlin 
states  have come forth\cite{Kane,TM}.

In this Letter, we propose a
 set-up consisting of two edge state
quasiparticle sources and two sinks, and the measurement of
current-current correlation for currents emerging
at the two sources and collected at the sinks. 
At equal times, the correlator
is found to depend only on average values of currents and
a factor $\cos{\pi\nu}$ coming from statistics. As a function of
time difference for when currents are correlated, it shows oscillations
with a period that depends on the fractional charge of the quasiparticle.

\begin{figure}[h]
\epsfxsize=2.5in
\centerline{\epsffile{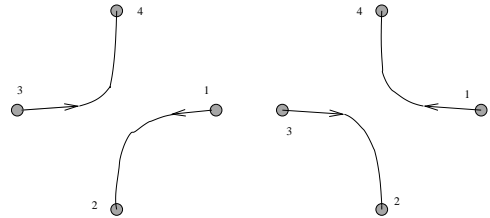}}
%\centerline{\psfig{exchange1.eps}}
\caption{Indistinguishable particles from sources `$1$' and `$3$' can reach
  sinks `$2$' and `$4$' by two possible processes whose probability 
amplitudes differ by a phase factor that depends on the statistics
of the particles.}
\label{scat}
\vspace{-0.2cm}
\end{figure}

Returning to the common wave function of Eq.(\ref{wvfn}), 
one can extract from it various properties, such as 
 filling in of available states, and $N$-point 
correlation functions. A quantity sensitive to statistics is
 the two-particle correlation function
\begin{equation}
g(r_1,r_2)  = N(N-1)\int dr_3\ldots dr_n
|\Psi(r_1,r_2...r_n)|^2, 
\label{twoptcorr}
\end{equation}
where `$N$' is the total number of particles.
In systems composed of a single species of non-interacting
particles, for fermions $g(r_1,r_2)$
necessarily drops to zero at $r_1=r_2$ and typically levels off to the
uncorrelated value for $r_1-r_2$ much greater than the mean particle
spacing. For uncondensed bosons, wave-function symmetrization allows 
$g(r_1,r_2)$ to reach twice its uncorrelated value. Our purpose here is to
study processes that enable the probing of anyonic systems
for {\it their} statistical information.
 Specifically, two-particle correlations are manifested in events
such as the ones shown in Fig.\ref{scat},
 where particles need not  scatter, but may merely be detected 
within a correlation region in time and space 
to feel the effect of statistics\cite{Gordon}. 

In fact, in the 1950's, Hanbury Brown and Twiss performed both astronomical
and table-top experiments to measure correlations in light intensities 
at two detectors, which strikingly reflected bosonic
statistics\cite{HBT,Gordon}. 
More 
 recently, laboratory experiments measuring 
analogous correlations in semi-conductors
 and in free space have brought out the fermionic statistics of
 electrons\cite{CS}. 
Turning our attention to Laughlin states, as anyons are only known to exist 
only within the strongly
correlated quantum Hall fluid, simplistic measurements requiring particles to
be detected outside the system would fail. Moreover, 
manipulations of quasiparticles within
the bulk are not feasible at present. However, it has been
suggested\cite{MatKane} and experimentally ascertained for $\nu=1/3$
states\cite{picc}, that weak tunneling of Laughlin quasiparticles between edge
states of a single Hall bar produces shot noise characteristic of particles
with quantized charge $\nu e$. 
Given the evidence for fractionally charged quasiparticles, we propose
a tunneling geometry that realizes the events depicted in Fig.\ref{scat}
for these particles. We show that appropriate 
current-current correlation measurements
in this geometry bring out the fractional statistics of the quasiparticles.

\begin{figure}[h]
\epsfxsize=2in
\centerline{\epsffile{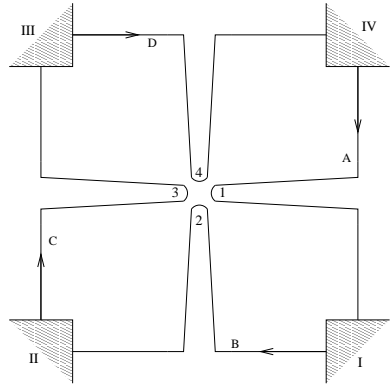}}
\caption{Set-up of Hall bar for measuring two-particle processes. Leads I-IV
  define edge states $A-D$. Pinching these edge-states together allows for
  inter-edge quasiparticle tunneling at points $1-4$.}
\label{cruciform}
\end{figure}

  Our proposed set-up is as shown in Fig.\ref{cruciform}. 
Four leads at the corners of the Hall bar define four edge 
states denoted by $\beta=A,B,C,D$. Low-energy excitations of the FQH system
correspond to long-wavelength density distortions of the edge. These
excitations can be described by the chiral Luttinger liquid
model\cite{Wen,Wenrev,Eduardo}, and thus each of the edge states is
 characterized by the Hamiltonian
\begin{equation}
H_0^{\beta} = \frac{1}{4\pi\nu}\int(\partial_x \phi_{\beta})^2 dx_{\beta},
\label{freeH}
\end{equation}
where the bosonic fields $\phi$ obey the commutation relations 
$[\phi_{\beta}(x), \phi_{\gamma}(x')]=i\pi\nu 
\rm{sign}(x-x')\delta_{\beta\gamma}$, and their gradients are
proportional to density distortions. 
Here, we have set the edge-state velocity to unity.
Gates allow for 
pinching the edge states close to one another\cite{picc} to form the cruciform 
pattern shown in Fig.\ref{cruciform}, thus enabling inter-edge 
quasiparticle tunneling. For each edge state $\beta$, we assume the 
tunneling to take place from points $x_j$ where $j=1,2,3,4$ for 
$\beta=A,B,C,D$ respectively. Here, we require that the region
formed by the tunneling points be comparable to the size of the 
quasiparticles. Unlike in the bulk, a second-quantized 
description of edge-state quasiparticles is relatively straightforward 
to formulate. We describe particles at 
the tunneling points  by the creation operators 
$\psi^{\dagger}_j = \kappa_j e^{-i \phi_{\beta}(x_j)}$, where the $\kappa$ 
denote Klein factors. The commutation relations for the bosonic fields of 
Eq.(\ref{freeH}) ensure that these quasiparticles, when exchanged with 
others residing on the same edge, exhibit the statistics of Eq.(\ref{wvfn}). 
The Klein factors ensure that they do so when exchanged with particles 
from neighboring edge states. We pick the convention
\begin{equation}
\psi^{\dagger}_j \psi^{\dagger}_k = 
e^{-i \pi\nu}\psi^{\dagger}_k \psi^{\dagger}_j,
\label{qpex}
\end{equation}
where $j<k$ for $j=1,2,3$, and $k=1$ for $j=4$.
Tunneling of these quasiparticles to 
neighboring edge states can be controlled by means of gate voltages.
It is described by the tunneling Hamiltonian
\begin{equation}
{\cal H}^{jk} = u_{jk}\psi^{\dagger}_j\psi_k^{ } + \rm{h.c.},
\label{tunham}
\end{equation}
where `h.c.' denotes Hermitian conjugation,  $\psi$'s are as  in 
Eq.(\ref{qpex}), and the $u$'s denote tunable bare tunneling strengths. 
As a variant of the Hanbury Brown-Twiss (HBT) experiment, we select
 the points $m=1,3$ as quasiparticle sources, and $n=2,4$ as sinks
by raising the potentials of the edge 
states $A$ and $C$ with respect to $B$ and $D$ by a voltage $V$. 
As all tunneling occurs in a fixed geometry, two-particle correlation
functions cannot  be studied as a function of spatial separation. However,
current-current correlations can be measured as a function of temporal
separation. These tunneling currents take the form
\begin{equation}
I_{mn}(t) = \frac{i e^*}{\hbar}(u_{mn}\psi^{\dagger}_m\psi_n 
e^{i \tilde{V}t} - \rm{h.c.}),
\label{tuncurr}
\end{equation}
where $e^*= \nu e$ is the charge of the quasiparticle
\cite{Mr.Bob,Eduardo,MatKane}, and 
$ \tilde{V}=e^*V/\hbar$. Edge states then carry measurable currents 
\begin{eqnarray}
I_I = \frac{\nu e^2}{h}V - I_{12} - I_{14} 
& ; & \hspace{0.1in} I_{II} = I_{12} + I_{32} \nonumber \\
I_{III}  =  \frac{\nu e^2}{h}V - I_{32} - I_{34} 
& ; & \hspace{0.1in} I_{IV} = I_{34} + I_{14},
\label{leadcurr}
\end{eqnarray} 
where in Fig.(\ref{cruciform}), $I_{\alpha}$  are currents going 
into leads `$\alpha$'.
The finite-temperature average values of these currents can be calculated 
using  non-equilibrium Keldysh techniques that treat tunneling 
perturbatively (see, for e.g., Ref.\cite{MatKane}.) 
To summarize the treatment, 
from Eq.(\ref{freeH}), we derive an action for each of the four edge states.
Away from the tunneling points $x_j$, the edge states are described
by free fields, whose form is explicitly obtained in terms
of the fields at the tunneling points using equations of motion. 
These free fields are integrated out to
obtain an effective action described by fields $\phi_j$ at 
points $x_j$. We then 
introduce a generating functional in terms of backwards and forwards 
real-time paths $\phi^{\pm}_j(t)$, which enables us to obtain 
expectation values. In equilibrium, the correlation function
$C_j\equiv\langle\varphi_j\varphi_j\rangle$ and response function
$R_j\equiv\langle\tilde{\varphi}_j\varphi_j\rangle$ , where
$\phi^{\pm}\equiv\varphi\pm\frac{1}{2}\tilde{\varphi}$, satisfy  
the fluctuation-dissipation theorem  
$C_j(\omega)=\coth(\hbar\omega/2kT)R_j(\omega)$. 
The effect of tunneling is treated perturbatively. 

To second order in tunneling, the technique described above
gives the following form for the average currents:
\begin{eqnarray}
& \langle I_{mn}(t)\rangle & =  \nu e 
\left(\frac{2 u_{mn}}{\hbar}\right)^2  \int d t_a \nonumber \\
&& \times\sin{\tilde{V}(t_a-t)}\sin{F(t_a-t)}e^{-f(t_a-t)},
\label{avcurr}
\end{eqnarray}
where $F(t)\equiv\nu\int d\omega \omega^{-1}\sin{\omega t}$ and 
$f(t)\equiv 2\nu\int d\omega \omega^{-1}\coth(\hbar\omega/2kT) 
(\sin{\omega t/2})^2$ originate from response and correlation 
functions respectively.
Evaluating the above gives for the differential
conductance contributions, $\frac{d I_{mn}}{d V} = 
V^{2\nu-2} {\cal G}(e^*V/k T)$, ${\cal G}(x\rightarrow\infty) 
\rightarrow \rm{const}.$, which increases with decreasing voltage. 
Thus, at low temperatures, which  is desirable for keeping thermal noise 
minimal, the voltage  $V$ must be held large compared to the bare 
tunneling strength for the perturbative treatment to remain valid.

We now analyze current correlations of particles emerging 
from the two sources detected at the two sinks.
In principle, a variety of current-current 
correlators contain  information on statistics. As was originally
 shown by Hanbury Brown and
Twiss, even particles from a single source can be distributed into two
detectors to exhibit statistical correlations\cite{HBT,Gordon}. 
Thus,  focusing on one set of 
source-drain edge states and measuring correlations between the 
transmitted and reflected currents \cite{Buttiker} along the source edge 
state, as was done for the integer quantum Hall system\cite{CS}, can give 
statistical information. Alternatively, the current-current 
correlations between 
currents from one source collected at  two different
 drains can be calculated\cite{Buttiker}, as has been done 
explicitly for the FQH set-up\cite{TM}. Here,
 we find that a clear-cut
signature of anyonic statistics comes from events shown in
Fig.\ref{scat}. To extract information on statistics
from these events, we propose the measurement of the 
following time-translation invariant current-current correlator:
\begin{equation}
{\cal C}(t-t') \equiv \langle\Delta I_{12}(t) \Delta I_{34}(t') + \Delta 
I_{14}(t) \Delta I_{32}(t')\rangle ,
\label{keycorr}
\end{equation}
where $\Delta I\equiv I-\langle I\rangle$. 

The correlator ${\cal C}$ can be obtained from three sets of
measurements. The first would measure correlations 
$\langle\Delta I_{II}(t)\Delta I_{IV}(t')\rangle$ for currents $I_{II}$
and $I_{IV}$ measured $\Delta t = t-t'$ apart. The other two sets 
of measurements would be performed in the absence of sources
`$1$' and `$3$', respectively,  realized by controlling 
the appropriate tunneling strengths $u_{mn}$ by means of gate 
voltages. It is important to note that these other sets of measurements
do not require a change in sample, but can be achieved merely
by applying the required gate voltages in a single sample. 
In each of these instances, currents into leads 
$II$ and $IV$ would have the form 
$\tilde{I}_{II}=I_{m2}$ and $\tilde{I}_{IV}=I_{m4}$ 
 with $m=3$ and $m=1$, respectively. Then, 
one could measure cross-correlations `$\tilde{C}$', 
 for current from one source held at the same potential $V$
as in the first case, into two drains, where 
$\tilde{C}_m(\Delta t)=\langle\Delta I_{m2}(t)\Delta I_{m4}(t')\rangle$,
with $m=3,1$ respectively, and $\Delta t = t-t'$.
 These correlations themselves carry
statistical information, but are complicated by the fact that 
the source edge states are endowed with their own dynamics.
Nevertheless, as seen in Ref.\cite{TM}, one can procure valuable 
information from them similar to that contained in our sought-after
correlator ${\cal C}(t-t')$ of Eq.\ref{keycorr}. This correlator
${\cal C}$ can now be obtained by subtracting  the contributions
of the latter two measurements from the first:
\begin{equation}
{\cal C}(\Delta t) = \langle\Delta I_{II}(t)\Delta I_{IV}(t')\rangle
- \tilde{C}_1(\Delta t)- \tilde{C}_3(\Delta t).
\end{equation}
In fact, Ref.\cite{Buttiker} proposes completely analogous sets of
measurements in a similar four point tunneling set-up in the integer
quantum Hall system, and there too, a correlator analogous to 
${\cal C}$ provides key information on statistics.

 The correlation can be evaluated in 
the perturbative Keldysh approach  outlined above. To lowest 
non-vanishing order, i.e., fourth order in tunneling, it takes the 
form 
\begin{equation}
{\cal C}(\Delta t)  =   \langle I_{12}(t)\rangle\langle I_{34}(t')\rangle + 
\langle I_{14}(t)\rangle\langle I_{32}(t')\rangle 
+ {\cal C}_{\diamondsuit}(\Delta t).
\label{keyencore}
\end{equation}
The function ${\cal C}_{\diamondsuit}(\Delta t)$ is the 
piece in the perturbation that connects all points $1-4$ of 
Fig. \ref{cruciform}, and thus contains information on the statistics. 
Explicitly, it is given by
\begin{eqnarray}
{\cal C}_{\diamondsuit}(\Delta t) & = & \cos{\pi\nu} (e^*)^2 \prod _{m,n} 
\frac{2u_{mn}}{\hbar} \int d t_a d t_b \times \nonumber
 \\ & & [\cos{\tilde{V}(t+ t' - t_a - t_b)} 
\cos{\tilde{F}} e^{-\tilde{f}}], 
\label{expc}
\end{eqnarray}
where $m=1,3$ and  $n=2,4$. 
Here, $\tilde{F}\equiv1/2\sum_{j=a,b}(F(t-t_j) + F(t'-t_j))$ and similarly 
$\tilde{f}$, involve the functions $F$ and $f$, which 
appear in Eq.(\ref{avcurr}). When the time difference $\Delta t$ is small,
 i.e. $\hbar/\Delta t\gg kT, e^* V$, one 
expects the correlations to be maximal\cite{Gordon}. In fact, in this 
limit and for uniform scattering $u_{mn} = u$ (which we assume from 
here on), upon evaluating Eq.(\ref{expc}), the current correlation defined 
in Eq.(\ref{keycorr}) reduces to the simple and suggestive form
\begin{equation}
{\cal C}(\Delta t \rightarrow 0) = 2 [1 + \cos{\pi\nu}] \langle I_{12}\rangle
 \langle I_{34}\rangle,
\label{maxcorr} 
\end{equation}
where the behavior of the average currents $\langle I\rangle$ is given in 
Eq.(\ref{avcurr}). This is consistent with the fermionic limit ${\cal C}(0) 
= 0$,  $\nu = 1$, which reflects the fact that two electrons cannot be in 
the same place simultaneously, and the bosonic limit of maximal ``bunching" 
for $\nu = 0$.
 
\begin{figure}[h]
\psfrag{C}{${\cal C}(\Delta t)$}
\psfrag{t}{$\Delta t$}
\epsfxsize=3in
\centerline{\epsffile{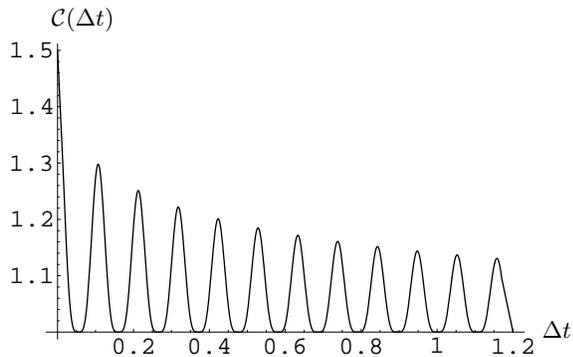}}
\vspace{0.15in}
\caption{Normalized correlation function of Eq.(\ref{keycorr}),
  $C(\Delta t)/(2\langle I_{12}\rangle\langle I_{34}\rangle)$, at
zero temperature, as a function of separation time
  $\Delta t$  for filling fraction $\nu=1/3$. Here we have chosen
  $\tilde{V}=60$ in dimensionless units.}
\vspace{0.15in}
\label{corrplot}
\end{figure}

 The function ${\cal C}(\Delta t)$ for finite $\Delta t$ carries telling 
information on two-particle correlations for edge-state quasiparticles.
At $T=0$, the integral of Eq.(\ref{expc})
can be evaluated using contour integration to give
\begin{eqnarray}
&&\frac{{\cal C}_{\diamondsuit}(\Delta t)}{2\cos{\pi\nu}}  = 
\left[\frac{4u^2e^*}{\hbar^2}\left(\frac{V}{\epsilon_0}\right)^{\!2\nu} 
\frac{e^{-e^*V/\epsilon_0}}{V} \frac{\pi \cos{\tilde{V}\Delta 
t/2}}{\Gamma(2\nu)\Gamma(1\!-\!2\nu)}\right]^2  \nonumber \\ 
 & & \ \ \times \left[\rm{Re}\left\{e^{-i\tilde{V}\Delta t/2} 
\int_0^{\infty}e^{-r} r^{-\nu} 
(r + i\tilde{V}\Delta t)^{-\nu}\right\}\right]^2,
\label{bigint}
\end{eqnarray}
where $\epsilon_0$, the excitation gap for the bulk Hall 
fluid, acts as a high-energy cut off. The resulting
behavior of the current-current correlations is shown in 
Fig.\ref{corrplot}.

The function ${\cal C}(\Delta t)$ contains 
 three different aspects of the edge state quasiparticles. First,
central to our problem and akin to 
the two-particle correlation function of Eq. \ref{twoptcorr}, 
it shows maximal statistical correlation at $\Delta t=0$. 
As shown by the factor of 
$\cos{\pi\nu}$ in Eq.(\ref{maxcorr}), 
it can either be smaller or larger than the uncorrelated value,
 depending on the (anti-)bunching 
nature of the quasiparticles. 
Second, ${\cal C}(\Delta t)$ reveals oscillations of period $h/e^*V$
and identifies $e^*=\nu e$ as the quantum of quasiparticle charge
that couples to the applied voltage. Note that for Laughlin
quasiparticles, statistics and charge are directly related to one
another, in that the phase factor acquired under exchange
may be interpreted as the Aharnov-Bohm term picked up
by the charged   quasiparticle\cite{Arovas,Wenrev}. However,
as the connection between charge and statistics is more complicated
for non-Laughlin states,\cite{Eduardo2,Wenrev}, in these cases, 
 ${\cal C}(\Delta t)$ becomes important in bearing information
on both aspects, distinct from one another. 
While the dependence on charge and statistics ought to hold
regardless of the effective theory used to describe the 
FQH system, the third feature reflects the chiral Luttinger
liquid description of the edge state
; at large separation time, ${\cal C}$ decays to the
uncorrelated value in the power-law form
${\cal C}_{\diamondsuit}(\Delta t)\sim |\tilde{V}\Delta t|^{-2\nu}$, 
 $|\tilde V \Delta t|\rightarrow\infty$, where the power-law behavior is
 characteristic of Luttinger liquids. [Note also that the
 static correlation function of Eq.(\ref{twoptcorr}) decays as
 $g(r_1,r_2)\sim|r_1-r_2|^{-2\nu}$ within a single edge state, in 
 contrast to the $|r_1-r_2|^{-2}$ decay appropriate to electrons in a
 1-dimensional Fermi liquid.] 
At finite temperatures, as seen from
Eq.(\ref{keyencore}) and Eq.(\ref{expc}), we expect the maximal
correlation function ${\cal C}(\Delta t=0)$ to cross over
to its uncorrelated value at temperatures $kT\approx e^*V$
\vspace*{-.1in} 

In conclusion, we have seen that the principle of extracting
information on statistics by means of two-point measurements
can be applied equally well to fractional particles realized
in current laboratory conditions as to the fermions and bosons 
found in Nature.
Characteristic correlations in the detection of
two particles at zero separation in space and time, their decay 
in space or time, and their oscillations over conjugate sets of
variables, be they  energy and time 
or position and momentum\cite{Gordon},
hitherto observed for fermions and bosons, are seen to be
manifest in processes involving anyons.
Given the current cutting-edge experimental developments
in quantum Hall physics, measurements on edge-state
quasiparticles such as the ones proposed here and in other 
work\cite{Kane,TM} ought to be within experimental reach,
and thus may provide signatures of fractional statistics for the
first time.

Warm and deep gratitude to E. Ardonne, G. Baym, M. P. A. Fisher, 
E. Fradkin, P. Goldbart and C. Kane for invaluable comments and 
discussions, and their caring support. This work was supported
by the grants NSF EIA01-21568, DOE DEFG02-96ER45434 and NSF PHY00-98353.

%\end{multicols}
\end{document}